\documentclass{article}
\usepackage[T1]{fontenc}

\usepackage{amsmath,amssymb}
\usepackage{microtype}
\usepackage{graphicx}
\usepackage{subcaption}
\usepackage{booktabs} 
\usepackage{xcolor}
\usepackage{multirow}               
\usepackage{array}                  
\newcommand{\R}{\mathbb{R}}
\newcommand{\D}{\mathcal{D}}
\newcommand{\M}{\mathcal{M}}

\newcommand{\animalA}{\mathsf{A}}
\newcommand{\animalB}{\mathsf{B}}
\DeclareMathOperator*{\median}{median}
\DeclareMathOperator{\corr}{\mathsf{Corr}}
\DeclareMathOperator{\rsa}{\mathsf{RSA}}

\DeclareMathOperator{\rdm}{\mathsf{RDM}}
\DeclareMathOperator{\train}{\mathsf{train}}
\DeclareMathOperator{\test}{\mathsf{test}}
\DeclareMathOperator{\trueA}{t^{\animalA}_{\train}}
\DeclareMathOperator{\trueB}{t^{\animalB}_{\train}}
\DeclareMathOperator{\trueBtest}{t^{\animalB}_{\test}}

\DeclareMathOperator{\trueAid}{t^{\animalA}}
\DeclareMathOperator{\trueBid}{t^{\animalB}}
\DeclareMathOperator{\sfAtrain}{s^{\animalA}_{1,\train}}
\DeclareMathOperator{\ssAtrain}{s^{\animalA}_{2,\train}}
\DeclareMathOperator{\sfBtrain}{s^{\animalB}_{1,\train}}
\DeclareMathOperator{\ssBtrain}{s^{\animalB}_{2,\train}}

\DeclareMathOperator{\sfBtest}{s^{\animalB}_{1,\test}}
\DeclareMathOperator{\ssBtest}{s^{\animalB}_{2,\test}}
\DeclareMathOperator{\sfAid}{s^{\animalA}_{1}}
\DeclareMathOperator{\ssAid}{s^{\animalA}_{2}}

\DeclareMathOperator{\sfBid}{s^{\animalB}_{1}}
\DeclareMathOperator{\ssBid}{s^{\animalB}_{2}}

\newcommand{\Mtrue}{\mathcal{M}_{\text{true}}}
\newcommand{\Mest}{\widehat{\mathcal{M}}_{\text{est}}}
\usepackage{listings}
\usepackage{xcolor}

\usepackage{hyperref}

\usepackage[accepted]{icml2025}
\makeatletter
\renewcommand{\icmlEqualContribution}{\textsuperscript{*}Equal contribution, authors listed alphabetically.}
\makeatother

\makeatletter
\renewcommand{\ICML@appearing}{Preprint.}

\def\@copyrightspace{%
  \@float{copyrightbox}[b]
  \begin{center}
  \setlength{\unitlength}{1pc}
  \begin{picture}(20,1.5)
    \put(0,0){\parbox[b]{19.75pc}{\small \Notice@String}}
  \end{picture}
  \end{center}
  \end@float
}

\makeatother

\usepackage{amsmath}
\usepackage{amssymb}
\usepackage{mathtools}
\usepackage{amsthm}

\usepackage[capitalize,noabbrev]{cleveref}

\theoremstyle{plain}
\newtheorem{theorem}{Theorem}[section]

\theoremstyle{definition}
\newtheorem{definition}[theorem]{Definition}

\theoremstyle{remark}

\usepackage[textsize=tiny]{todonotes}

\icmltitlerunning{NeuroAI Turing Test}

\begin{document}

\twocolumn[
\icmltitle{Brain-Model Evaluations Need the NeuroAI Turing Test}

\icmlsetsymbol{equal}{*}

\begin{icmlauthorlist}
  \icmlauthor{Jenelle Feather}{equal,yyy}
  \icmlauthor{Meenakshi Khosla}{equal,comp}
  \icmlauthor{N. Apurva Ratan Murty}{equal,sch}
  \icmlauthor{Aran Nayebi}{equal,cmu}
\end{icmlauthorlist}

\icmlaffiliation{yyy}{Flatiron Institute, New York, USA}
\icmlaffiliation{comp}{University of California San Diego, La Jolla, USA}
\icmlaffiliation{sch}{Georgia Institute of Technology, Atlanta, USA}
\icmlaffiliation{cmu}{Carnegie Mellon University, Pittsburgh, USA}

\icmlcorrespondingauthor{Jenelle Feather}{jfeather@flatironinstitute.org}
\icmlcorrespondingauthor{Meenakshi Khosla}{mkhosla@ucsd.edu}
\icmlcorrespondingauthor{N. Apurva Ratan Murty}{ratan@gatech.edu}
\icmlcorrespondingauthor{Aran Nayebi}{anayebi@cs.cmu.edu}

\icmlkeywords{NeuroAI, Turing Test, Machine Learning}

\vskip 0.3in
]

\begin{abstract}
What makes an artificial system a good model of intelligence?
The classical test proposed by Alan Turing focuses on behavior, requiring that an artificial agent's behavior be indistinguishable from that of a human.
While behavioral similarity provides a strong starting point, two systems with very different internal representations can produce the same outputs.
Thus, in modeling biological intelligence, the field of NeuroAI often aims to go beyond behavioral similarity and achieve representational convergence between a model's activations and the measured activity of a biological system.
This position paper argues that the standard definition of the Turing Test is incomplete for NeuroAI, and proposes a stronger framework called the ``NeuroAI Turing Test,'' a benchmark that extends beyond behavior alone and \emph{additionally} requires models to produce internal neural representations that are empirically indistinguishable from those of a brain up to measured individual variability, i.e. the differences between a computational model and the brain is no more than the difference between one brain and another brain.
While the brain is not necessarily the ceiling of intelligence, it remains the only universally agreed-upon example, making it a natural reference point for evaluating computational models.
By proposing this framework, we aim to shift the discourse from loosely defined notions of brain inspiration to a systematic and testable standard centered on both behavior and internal representations, providing a clear benchmark for neuroscientific modeling and AI development.
\vspace{10pt}
\end{abstract}
---------------------------------------------
\printAffiliationsAndNotice{\icmlEqualContribution}

\section{Introduction}
\label{sec:intro}

\begin{figure}[t]
    \centering
    \includegraphics[width=\linewidth]{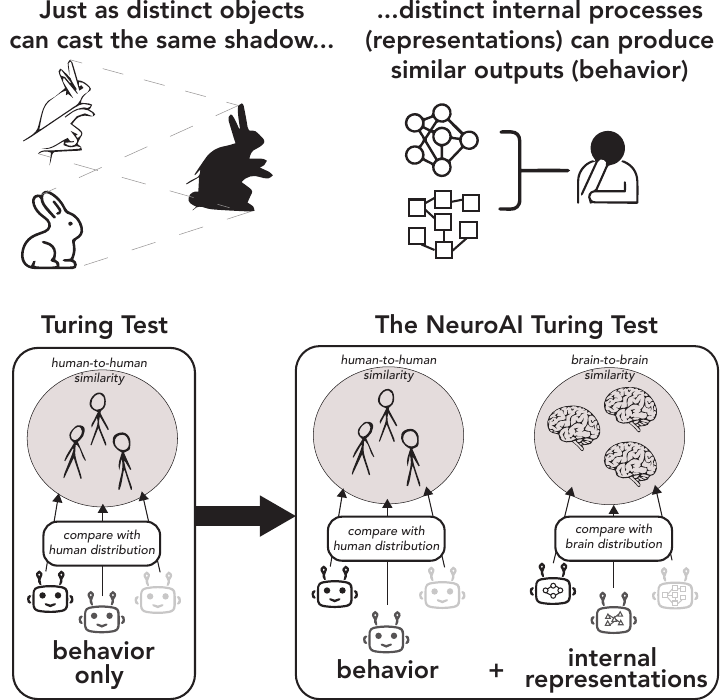}
    \vspace{-20pt}
    \caption{The complete NeuroAI Turing Test reflects the similarity of both the behavior of an artificial system and the similarity of the internal representations.}
    \label{fig:neuroaiTuringTest}
    \vspace{-15pt}
\end{figure}

Humanity is in the midst of an intense pursuit to understand and replicate intelligence in artificial systems.
But the AI community (computer scientists scaling up AI algorithms) and the NeuroAI community (computational cognitive neuroscientists leveraging AI systems to build models of the brain) seem to be approaching the challenge of intelligence from fundamentally divergent perspectives.
AI researchers have primarily focused on developing systems that exhibit intelligent behavior, a tradition rooted in Alan Turing’s seminal idea of the Turing Test~\citep{turing1950computing}. 
The focus on achieving behavioral similarity to humans has undoubtedly propelled AI forward, but the goal is rarely to study and understand how the internal representations of AI systems generate behavior.
In contrast, NeuroAI aims to model the computational principles underlying intelligence in biological systems, using the brain—the only example of intelligence we universally recognize and agree upon—as a reference point.
Evaluations of NeuroAI models often involve collecting brain data (e.g., neural spike trains or fMRI data) and comparing the similarity of the evoked representations with those measured from an artificial system. 
While the brain serves as a natural ground-truth for intelligence, providing a benchmark for any system aspiring to match or exceed it, we have yet to settle on the core algorithmic principles that give rise to many of its intelligent behaviors. 
Meeting this threshold is a necessary step before we can surpass it, yet despite significant advancements in our ability to collect and analyze neural data, NeuroAI has not established a clear set of criteria for when a model would successfully be considered ``brain-like.''

Here we propose the NeuroAI Turing Test to address these challenges. \textbf{Our central position is that brain-model evaluations in computational neuroscience urgently need a NeuroAI Turing Test that assesses both behavior and internal representations to the limit of carefully recorded biological measurements (the ceiling).} 

The NeuroAI Turing Test extends Turing’s original evaluation framework by integrating representational convergence of the internal features into the evaluation criteria (the integrative benchmarking condition \cite{schrimpf2020integrative}). 
It also mandates that an artificial model’s internal neural activations must be empirically indistinguishable from those observed in biological brains, but within the bounds of natural inter-individual variability (a measure of success that goes beyond integrative benchmarking). 

We propose that the NeuroAI Turing Test become the standard in NeuroAI, towards the development of truly brain-like artificial intelligence. Our proposition is timely and necessary. As AI systems become increasingly sophisticated, the demand for models that genuinely reflect human intelligence will intensify. The NeuroAI Turing Test offers a testable standard that brings together the behavioral focus of AI with the mechanistic insights of neuroscience.

\section{Why do we need a distinct test for NeuroAI?}
\label{sec: need}

\textbf{We think that the NeuroAI Turing Test is imperative for overcoming fundamental limitations inherent in both the classical Turing Test and the prevailing practices within NeuroAI.} 

The traditional Turing Test and recent proposed extensions 
\citep{zador2023catalyzing}, focus extensively on behavioral indistinguishability between biological organisms and machines i.e. the outputs.
However, testing the output behavior alone is insufficient because many possible internal processes can produce identical behaviors through entirely different computational mechanisms (see Fig.~\ref{fig:neuroaiTuringTest} top).

NeuroAI on the other hand, places importance on faithfully modeling both the human brain and behavior (cognition). But even within the NeuroAI community,  multiple definitions exist for what constitutes a ``good'' model.
In addition, the methodologies for comparing models with brains are fragmented, and rely on evaluation measures (metrics) that offer only relative comparisons between models without establishing any specific consideration for a definitive standard of success (more on this to follow).

The NeuroAI Turing Test directly addresses these shortcomings by establishing rigorous criteria for replicating behavior and internal mechanisms (biological brains) benchmarked against the variability observed between individual humans (or animals, in the case of many neural recordings).
To assess internal similarity, we focus on measures of \textit{representational convergence} which capture similarity at Marr's ``algorithmic'' level, in contrast to the original Turing Test which evaluates similarity at the ``computational'' level \cite{marr2010vision}. Note however that the NeuroAI Turing Test framework is also sufficiently general to also be applied to models that attempt to capture ``implementation'' level constraints. 

Together, the NeuroAI Turing Test sets a higher bar for scientific rigor and empirical validation for brain models. Our goal is to empower the AI and Neuroscience communities with the tools and methods to identify superficial imitations and to strive for neuroscientific accuracy with brains. We think this shift is crucial for steering the co-evolution of AI and neuroscience toward new insights and the development of truly robust, brain-inspired systems. Without a rigorous standard, AI risks perpetuating models that are impressive in performance yet fundamentally disconnected from the biological intelligence they seek to emulate. The NeuroAI Turing Test is a necessary evolution.

\begin{figure}
    \centering
    \includegraphics[width=\columnwidth]{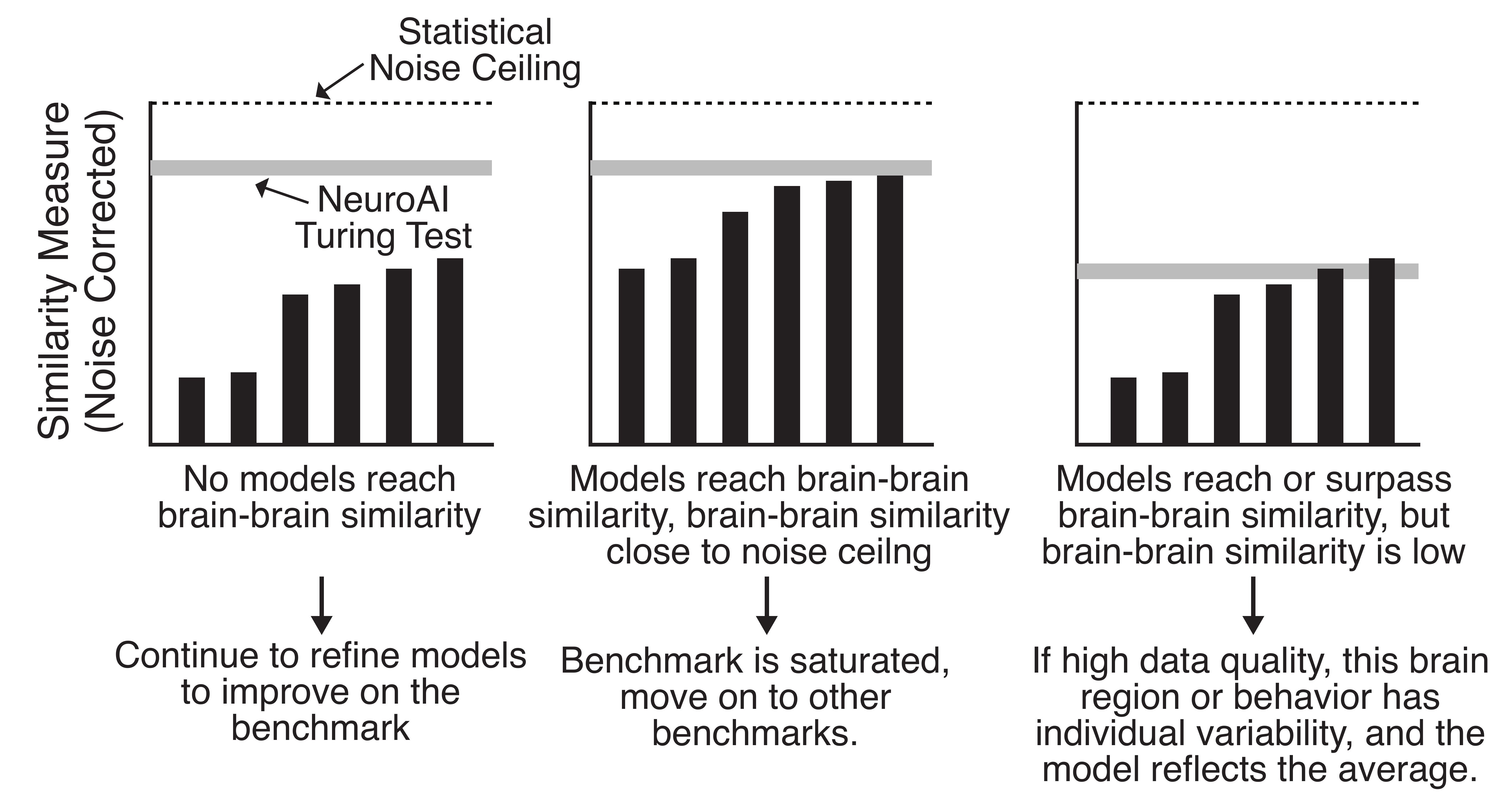}
    \vspace{-10pt}
    \caption{Possible outcomes of a NeuroAI Turing Test (brain-brain similarity). Each bar is for a different model. The model similarity measure and the brain-brain similarity (NeuroAI Turing Test) are both corrected by the square root of the product of the internal and mapping consistencies, constituting the ``Statistical Noise Ceiling'' (see Appendix~\ref{sec:methods-interanimal} for details). Different interpretations arise from the relationship between the model similarity and the distribution of the brain-brain similarity. Researchers should report both values to ensure that a benchmark is not saturated according to brain-brain similarity. Although this figure focuses on the alignment of internal representations, similar comparisons should be reported for behavioral tests.
    }
    \label{fig:possible_outcomes}
    \vspace{-15pt}
\end{figure}

\begin{figure}[htb]
    \centering
    \includegraphics[width=\columnwidth]{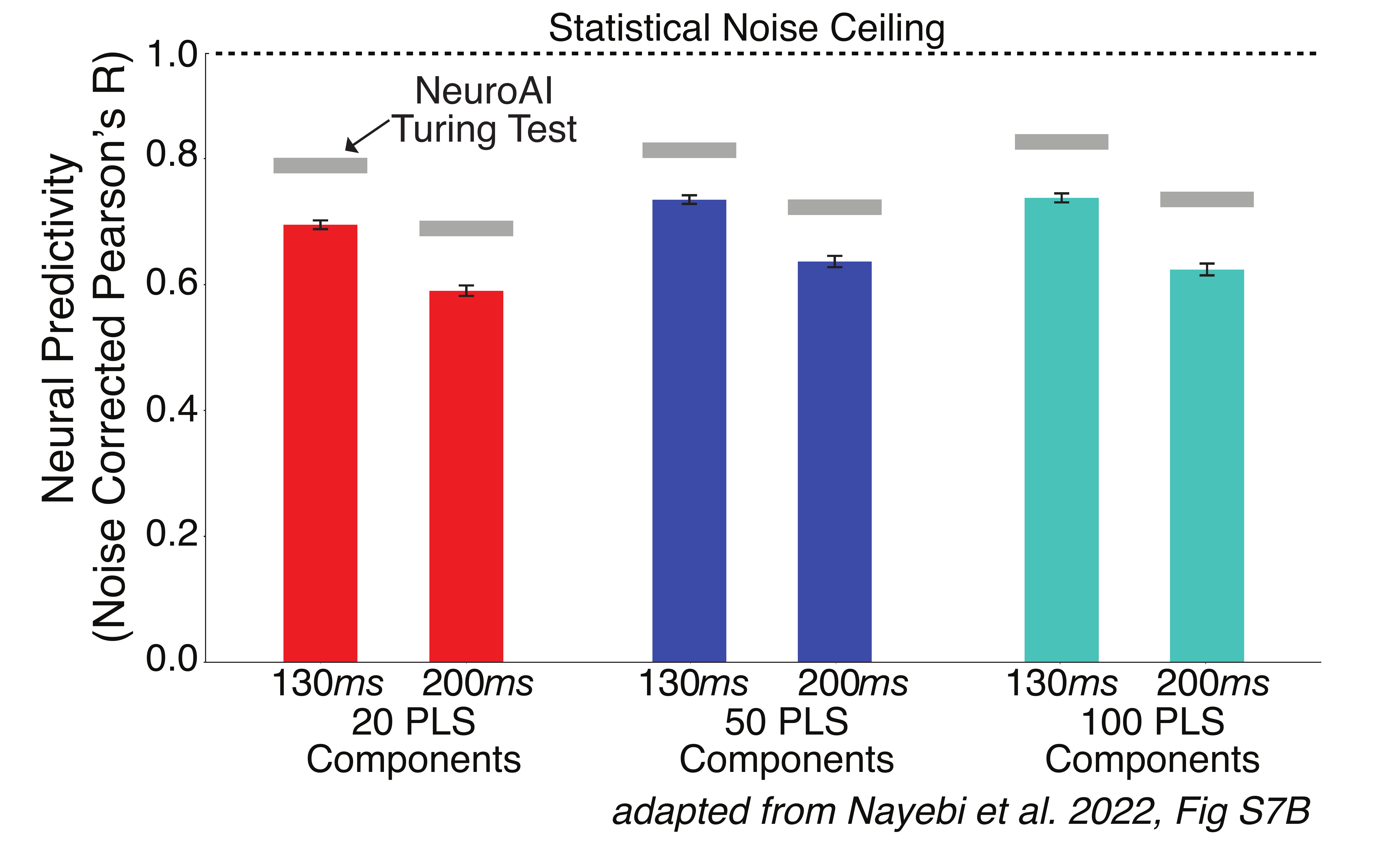}
    \vspace{-10pt}
    \caption{The NeuroAI Turing Test on the classic HvM dataset~\citep{Majaj2015} with different metrics. Neural predictivity is shown for a ResNet-18 inspired feedforward network from~\citet{nayebi2022}. The linear mapping was performed on multiple timepoints with different numbers of PLS components, and each dataset and measure has a separate value for the NeuroAI Turing Test. These data suggest that at least for HvM, we have reasonably saturated this benchmark and should choose other ones for primate vision.}
    \label{fig:hvm_analysis}
    \vspace{-10pt}
\end{figure}

\textbf{Motivation for standardization.}
Often, measures of brain-brain similarity are relegated to the supplement of papers. These numbers are typically below the noise ceiling obtained from data splits within the same participants, and the amount of variance yet to explain relative to the noise-ceiling is emphasized as progress to be made with future models.
There is also a lack of agreement about what value on a particular dataset and metric would suggest that the similarity measure has been saturated, reflected by a diverse set of ceiling measurements in common benchmarks \cite{Schrimpf2018, ratan2021computational, conwell2022can}. 
Our NeuroAI Turing test provides practical guidance for future reporting of benchmarks: in addition to noise correcting for variability due to internal stochasticity and measurement noise, we must also report brain-brain similarity values (Fig.~\ref{fig:possible_outcomes}).
And when we do, we find on ``classic'' standard benchmarks (such as for primate object recognition), that current models are closer to saturating them than we thought before (Fig. ~\ref{fig:hvm_analysis}), suggesting a greater need for \emph{new} benchmarks, rather than pushing further on the existing ones to yield incremental improvement.

\section{Defining the NeuroAI Turing Test}
\label{sec:def-neuroai-turing}
In the standard Turing Test, Alan Turing reframed the question ``Can machines think?'' into ``Can machines produce behavior indistinguishable from a human?'' 
Similarly, in the NeuroAI Turing Test, we extend this idea by requiring machines not only to produce behavior that aligns with human capabilities but also to generate \emph{internal} representations that are indistinguishable from those recorded from human (or animal) brains.

\textbf{Setup.} Let $\D \in \mathbb{R}^{C \times T \times N}$, where $C=\text{(\# conditions)},\;T=\text{(\# timepoints)},\;N=\text{(\# outputs)}$, be a dataset of neural (e.g., neuron spike counts or fMRI voxel responses) or behavioral responses from a set of organisms (animals or human subjects) $\mathcal{O}(\mathcal{D})$.
Let $\{X_i\}_{i \in \mathcal{O}(\mathcal{D})}$ be the measurements from those organisms.
In this framework, the time dimension is optional, as one can examine time averages, but at minimum conditions and outputs must be obtained from the brain or behavior.
Let $X_m$ be the corresponding measurements from a model (e.g. unit activations or behavioral output from a neural network) under the same conditions.
Note that the model can either be embodied~\citep{zador2023catalyzing,pak2023newborn} or not, that is a flexible choice.
Let $\M : \mathbb{R}^{C \times T \times N} \to \R$ be a metric on the space of these representations.

\textbf{Distances.}
Define the \emph{inter-organism} distance set:
\begin{equation*}
    \Delta_{\text{organism}} \;=\; 
    \bigl\{\, \mathcal{M}(X_i, \, X_j) 
    \;:\; i,j \in \mathcal{O}(\mathcal{D}),\, i \neq j \bigr\}.
\end{equation*}
Next, define the \emph{model-organism} distance set:
\begin{equation*}
    \Delta_{\text{model}} \;=\;
    \bigl\{\, \mathcal{M}(X_m, \, X_i) 
    \;:\; i \in \mathcal{O}(\mathcal{D}) \bigr\}.
\end{equation*}

\textbf{Hypothesis Testing.}
Next, choose a significance level $\alpha \in (0,1)$ of convergence.
Select a distribution-free two-sample test $T$ (e.g., rank-sum, permutation, or KS) to compare $\Delta_{\text{organism}}$ and $\Delta_{\text{model}}$.
Let
\begin{equation*}
\begin{split}
& H_0:\ \Delta_{\text{model}} \text{ and } \Delta_{\text{organism}} 
    \text{ come from the same distribution,}\\
& H_1:\ \Delta_{\text{model}} \text{ and } \Delta_{\text{organism}} 
    \text{ differ systematically } \\
& \quad \text{(e.g., model-organism distances are larger).}
\end{split}
\end{equation*}

\begin{definition}[Convergence in Distribution.]\label{def:rep-conv}
We say the model's representation \emph{converges in distribution} to that of the organisms in $\mathcal{O}(\mathcal{D})$ (at level $\alpha$) if:
\begin{enumerate}
    \item Under the chosen test $T$, we \emph{fail} to reject $H_0$ at the $\alpha$ level (i.e., $p \ge \alpha$),
    \item A chosen statistic (e.g., difference in means or medians) indicates that model-organism distances are not systematically larger (or otherwise different) than the inter-organism distances.
\end{enumerate}
\end{definition}

Note that our definition of the NeuroAI Turing test is applicable to any measure of similarity used to compare a computational model with a biological system.
The choice of this measure $\mathcal{M}$ is determined by the user, but we highlight some considerations here(for more discussion see Appendix \S\ref{ss:notions-metrics}, Table~\ref{tab:metrics}).
Common choices of mapping function and metric include linear regression (ridge, PLS), and RSA; however, it is recommended to restrict oneself to the ``sharpest'' transform class depending on the scientific question at hand~\citep{thobani2024inter}, to be in a position to most stringently separate models.
Note that for common choices of metric $\mathcal{M}$, such as Pearson correlation, RSA, and especially any metric that satisfies transitive closure~\citep{williams2021generalized}, one will need to correct by the square root of the product of the mapping consistency and internal consistency of the units, in order to properly approximate the true value of $\mathcal{M}$ in the limit of infinite trials.
We provide a heuristic derivation of why this correction needs to be encoded in $\mathcal{M}$ in Appendix \S\ref{sec:methods-interanimal}.
Finally, in the case of regressions, which return a per-unit consistency value, if we want $\mathcal{M}$ to be a real number per subject, we recommend taking the median across the population to avoid sensitivity to outliers, following \citet{yamins2014performance} and many other subsequent works.
One can also plot these consistencies per unit if they wish, rather than collapsing to a single number.

While this paper centers on our proposed definition of the NeuroAI Turing Test, models that explicitly posit a mechanism of variability could be held to a higher standard: the variability among model instances should mirror that of real animals (see Appendix \S\ref{ss:notions-variability}). This type of variability may be especially important when studying higher-order cognitive functions, where subjects often take different \emph{strategies}. For this work, however, we focus on testing models based on the convergence of distribution between $\Delta_{\text{model}}$ and $\Delta_{\text{organism}}$. Even in situations where there is a large amount of inter-subject variability, one can learn about commonalities in the representations by focusing on metrics that identify common features \emph{across} different individuals (see~\citet{nayebi2021explaining} for an example of this approach).

\section{Current State of the NeuroAI Turing Test}
\label{sec:current-state}
With the formalized  NeuroAI Turing test, we detail how current work on behavioral and representational convergence fits into this framing. 

\textbf{Behavioral alignment alone is not sufficient for representational convergence between models and brains.}
Let's consider a model that passes the behavioral Turing Test. Would this model also pass a representational level test for brains? 
Some researchers have argued that several high-capacity neural networks are converging to universal ``platonic representations''  \citep{huhposition}. This idea is closely related to notions of efficient coding \cite{barlow1961possible, simoncelli2001natural} in neuroscience. But note converging to the \textit{optimal} code still critically depends on strong implementational constraints. And the constraints in biology are quite different than those in artificial systems. 
Some work has suggested that including biological constraints in the architecture or training environment of models leads to better representational alignment \cite{dapello2020simulating, dapello2021neural,anonymous2025toponets}. It remains an open question whether (and which) of these biological constraints critically shape the representation, such that a high-capacity system without such constraints would yield a different optimal solution. Thus, we argue that testing behavior alone is not \textit{sufficient}, and we need models that also achieve representational convergence with the brain. 

\begin{figure*}[htb]
    \centering
    \includegraphics[width=\linewidth]{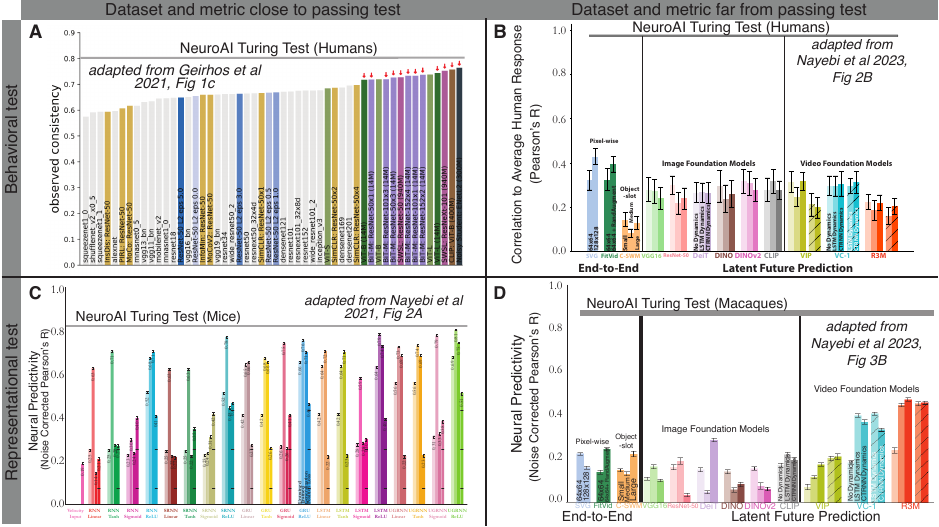}
    \vspace{-20pt}
    \caption{Examples of previous studies using behavioral and representational similarity tests. Artificial models have reached biological behavior and representational similarity in some datasets but not others.}
    \label{fig:previous_examples}
    \vspace{-15pt}
\end{figure*}

\textbf{Behavioral alignment is necessary for a complete model of the brain to pass the NeuroAI Turing test.}
Is passing a behavioral test \textit{necessary} for passing our NeuroAI Turing test? 
If the goal is to model the entire organism, passing a behavioral test is necessary. 
Take a simple deterministic example where $f(x)$ is an internal representation and $g(x)$ outputs behavior: if
$g(f(x)) \ne g'(f'(x))$ then we know that $g'\ne g$ and/or $f' \ne f$. 
Models in recent years have made immense progress at solving complex behavioral tasks, including recent work showing increasing convergence in domains that previously had large gaps~\citep{geirhos2021partial} (Fig.~\ref{fig:previous_examples}A shows one metric from this work). 
However, even in well-studied domains such as vision, these models are ``not yet adequate computational models of human core object recognition behavior'' \citep{wichmann2023deep}.
There are many instances of models failing to reach human-like performance on behavioral measures of brain-model alignment~\citep{feather2019metamers,feather2023model, hermann2020origins,baker2022deep,bowers2023deep}.
The gap is especially pronounced for complex dynamic stimuli, such as future prediction \citep[Fig 2A]{nayebi2023neural} (key result reproduced in Fig.~\ref{fig:previous_examples}B).

We emphasize that these proposed behavioral measures (and non-correlational controlled experiments) are compatible with our NeuroAI Turing Test, as they amount to a choice of metric $\mathcal{M}$.
However, our test goes beyond merely specifying a metric---it establishes a well-defined inter-subject behavioral consistency ceiling against which models can be evaluated. 
Without such a measure, one risks drawing incorrect conclusions, such as the claim that model-IT predictivity is disproportionately driven by background-processing~\citep{malhotra2024predicting}.
This misunderstanding arises from conflating ``core object recognition'' with category-only processing, despite well-established evidence that IT itself encodes category-orthogonal features~\citep{Hong2016}. 
It is therefore critical to consider what is already accounted for in the neural data relative to models; failing to do so in this case overlooks background-related processing in IT, leading to misleading model comparisons.

Further, models achieving ``super-human'' performance on tasks are generally not considered brain-like, as their behavior could easily be distinguished from humans by looking at these abilities (for instance, in \citet{zhang2022human} the authors state that ``advanced scientific topics'' can serve as a way to discern human and AI agents, as the AI does not find the specialized jargon difficult to discuss). With the lack of clear behavioral alignment between brains and machines, we can conclude that we have yet to achieve a foundation model of the whole brain that would pass our NeuroAI Turing Test. 

\textbf{NeuroAI Turing test for individual brain modules.} Currently in neuroscience, most researchers focus on building models for a subset of neural responses (for instance, one particular visual area), rather than the entirety of the brain. In these cases, even if a model's behavior is not aligned with humans, it might encode a good representation for an intermediate brain module. While our long-term goal should be for models that capture all neural components, targeting modules of the brain is one stepping stone for building a larger-scale system. In some domains and datasets, models have begun to come close to or achieve representational-level convergence with biological measurements. For instance, in common vision benchmarks from macaque inferior temporal (IT) cortex \citep{Majaj2015}, and human fMRI data on naturalistic stimuli,  neural predictivity seems to be at 90\% of the inter-animal variability ~\citep{nayebi2022, conwell2022can, ratan2021computational}. 
In rodent (mouse and rat) medial entorhinal cortex, models closely approach the inter-animal consistency measure~\citep[Fig. 2A]{nayebi2021explaining} (replotted in Fig.~\ref{fig:previous_examples}C), and in mouse visual cortex, models also achieve nearly 90\% of the inter-animal variability~\citep[Figs. 2A, S2, and S3]{nayebi2021unsupervised}. In cases where predictions on the dataset have been nearly saturated with respect to the inter-animal baseline, it is critical to examine the dataset and metrics that have been used to gain insight for future experiments. Other benchmarks show a clear gap between the best models and inter-animal consistency.  
In primate dorsomedial frontal cortex, the neural predictions from the best models are far below that obtained from inter-animal consistency \citep[Fig 3B]{nayebi2023neural} (replotted in Fig.~\ref{fig:previous_examples}D), and measures of RSA with auditory fMRI data show a clear different from a between-participant ceiling \citep[Fig 2E]{tuckute2023many}. 

\section{Alternate Views}
\label{sec:against}
In this section, we discuss alternate views to our proposed NeuroAI Turing Test.
We group the main arguments against it by subsection:
\subsection{Alternate Notions of ``Brain-Likeness''}
\label{ss:against-likeness}
The past century of neuroscience has had many definitions of what a brain ``does'', thereby providing an implicit definition for what might constitute a ``brain-like'' model (cf. Table~\ref{tab:brain_likeness} and Appendix \S\ref{sec:notions}). 
For example, the most notable of these include classic ideas of predictive coding~\citep{rao1999predictive}, sparse coding~\citep{olshausen1996emergence}, energy efficiency~\citep{laughlin2001energy}, and redundancy reduction~\citep{barlow1961possible}.

While these notions are valuable, they do not inherently define a level of convergence that researchers should aim to achieve for a model of the brain. 
Our requirement of distributional convergence, in contrast, remains agnostic to any single (and potentially subjective) principle of how brain-likeness \emph{ought} to be defined \emph{a priori}. Instead, the metric $\mathcal{M}$ encodes the user's preference for which features of brain-likeness they wish to align with, depending on their specific scientific question. Importantly, one is not restricted to a single choice of metric; each metric induces a distinct NeuroAI Turing Test, capturing a different dimension of ``brain-likeness'' to investigate. This flexibility allows our framework to encompass and generalize past notions of brain-likeness by requiring that models are distributionally similar to brains under the chosen metric.

Other work has focused explicitly on the biological plausibility of models as the gold standard for `brain-likeness'. But biological plausibility alone has not yet led to systems that capture complex behavior.
Spiking neural networks, for example, while more biologically plausible than a typical neural network architecture, often underperform in predicting neural responses or behavior. On the other hand, the constraints imposed by biologically inspired components have led to models that are closer to human behavior~\citep{saddler2021deep} or better at predicting neural responses~\citep{dapello2020simulating}.
These examples highlight that behavioral similarity, representational convergence, and biological plausibility are complementary yet distinct goals. Here, we focus on both behavior and representational convergence, as the goal of NeuroAI is typically not to \emph{emulate} the brain at every level of biological detail, but rather to understand and abstract the biological details necessary for intelligent behaviors, by communicating the evolutionary constraints of brain function through predictive model tasks and architecture classes.

\subsection{Beyond ``Brain-likeness'': Predictivity, Interpretability, and Practical relevance\label{sec:metric_interpretability}}
\textbf{Predictivity vs. Interpretability.}
There is a perceived tension between approaches that optimize for neural predictivity and approaches that seek interpretable explanations for neural phenomena.  Predictive approaches prioritize obtaining quantitatively accurate replications of neural activity, while interpretability approaches seek to uncover human-explainable insights into the causes of neural phenomena.
Handcrafting mechanistic explanations in neuroscience has always been challenging, a situation analogous to the ``bitter lesson''~\citep{sutton2019bitter} in AI. 

Interpretability, however, is an inherently subjective measure, often in the eye of the beholder. Previous approaches that prioritized \emph{particular} mathematically interpretable features consistently fell short in accurately predicting large-scale neural activity. This is very likely because neural populations are heterogeneous and inherent difficult-to-interpret and hand-engineer. Notable examples of these failures across species and brain areas include Gabors and macaque V1~\citep{cadena2019deep}, HMAX~\citep{riesenhuber1999hierarchical} and macaque V4/IT shape parameterization methods ~\citep{yamins2014performance, pasupathy2001shape, pasupathy2002population, yamane2008neural}, and grid-cell-only models and rodent MEC~\citep{nayebi2021explaining}.

Critically, the predictive modeling strategy is \textit{not} antithetical to the NeuroAI Turing Test or those aiming to uncover an interpretable understanding of neural responses.  It is  much easier to seek interpretable explanations for scientific phenomena in a task-optimized \emph{in silico} system, within which we have perfect access to every computation, than biological brains which are significantly more noisy, and limited-access (see, e.g.~\citep{mcintosh2016deep,tanaka2019deep,maheswaranathan2023interpreting, tanaka2019deep, nayebi2022goal, ratan2021computational, khosla2022highly}, for how this can be achieved in artificial neural networks (ANNs) relevant to neurobiology). In fact NeuroAI Turing Test, strongly supports attempts to extract interpretable explanations and mechanistic understanding of neurobiological phenomena. It makes this enterprise more rigorous by enforcing a critical constraint: all benchmarks, including mechanistic explanations, must be evaluated against a carefully measured inter-animal ceiling.

\textbf{Beyond Neuroscience: AI Safety.} Even beyond neuroscience, the NeuroAI Turing Test is in alignment with AI mechanistic interpretability and safety efforts that have been trying to make sense of the model internals ~\citep{olah2020zoom,nanda2023comprehensive,rager2025auditing}.
We are suggesting a step further by enforcing that what is found mechanistically should be \emph{replicable} across individuals; otherwise, it risks being a spurious, idiosyncratic observation.
This convergence not only bridges neuroscience and AI safety but also paves the way for interdisciplinary methodologies that enhance our understanding of complex systems.

\textbf{Practical applications: Brain-Computer Interfaces (BCIs).}
Accurate neural prediction is also intrinsically valuable for BCIs.
``Digital twins'' of neural responses can be manipulated to test interventions~\citep{lee2020topographic,schrimpf2021topographic,pak2023newborn,anonymous2025toponets} or produce stimuli that optimally drive biological neurons~\citep{bashivan2019neural,walker2019inception, tuckute2024driving}. These results show that once a model reliably predicts a system, it can also \emph{control} the system---a principle seen in many scientific advances. For example, advances in quantum mechanics laid the groundwork for nuclear physics, which ultimately enabled the development of nuclear reactors to control subatomic processes. For BCIs, matching the \emph{individual} may be ultimately more important than matching the population distribution, but achieving reliable alignment at a population level is a first step before tailoring person-specific solutions. If the broader distribution remains elusive, personalized alignment becomes even more difficult.

\subsection{Against Integrative Benchmarking More Generally: ``Goodharting''}
\label{ss:against-goodharting}
A common criticism of measurement and benchmarking strategies is their susceptibility to Goodhart’s law~\citep{goodhart1975problems}: ``When a measure becomes a target, it ceases to be a good measure.''
When a single benchmark becomes the primary objective, researchers may optimize for it at the expense of broader goals it was meant to capture. 
This issue is not unique to our proposed NeuroAI Turing Test but applies to integrative benchmarking more generally.

We address the more general criticisms against integrative benchmarking here, but note that the NeuroAI Turing Test already avoids this pitfall by not being tied to any \emph{one} metric. 
Instead, it is \emph{parametrized} by a choice of metric $\mathcal{M}$, with a flexible threshold $\alpha$ that allows for distributional rather than exact equivalence. 
This adaptability ensures that the NeuroAI Turing Test is not a static benchmark that can be trivially over-optimized but a framework that evolves with different contexts.

It has also been argued that some metrics might encode biases to ``pick out'' outliers in a high-dimensional space (a consequence of the ``Ugly Duckling Theorem''~\citep{watanabe1969knowing}, the basis for the ``No Free Lunch Theorem''~\citep{mitchell1980need,wolpert1997no}).
While this is theoretically possible, one should note that even using metrics alone \emph{independently} of the NeuroAI Turing Test, has not resulted in ``Goodharting''. Experimental results across domains and datasets paint a consistent picture: the model-to-brain matching metrics have not unilaterally gone up, but have instead either plateaued or even gone down~\citep{schrimpf2020integrative, conwell2022can, tuckute2023many, soni2024conclusions, linsley2023performanceoptimizeddeepneuralnetworks}.
Simple model manipulations do not lead to a perfect match~\citep{riedel2022bag}.
Rather, the largest advances in neural and behavioral predictivity have come from significant AI advances, such as ImageNet-categorization-optimized CNNs over everything before it (e.g. handcrafted SIFT features, HMAX, as in \citet{yamins2014performance}), deeper CNNs like ResNets over AlexNet~\citep{schrimpf2020integrative,Kar2021}, or the advent of performant contrastive, self-supervised learning objectives~\citep{Zhuang2021,nayebi2021unsupervised,konkle2022self, yerxacontrastive}.
In recent analyses, pretraining dataset scaling has shown to be related to improved behavioral alignment \cite{geirhos2021partial}, but this does not appear sufficient for neural alignment~\citep{gokce2024scaling}.
This indicates that current models may not be sufficient and innovations in inductive biases are likely needed for improved neural alignment. This is particularly true in domains such as 3D vision and embodied, physical intelligence~\citep{hermann2023humanlike}, where large-scale data collection is feasible but yields significantly smaller performance gains compared to language, as seen with existing Vision Transformers~\citep{bear2021physion,nayebi2023neural,bear2023unifying,vc2023}.

Furthermore, even \emph{despite} differences between regressed and non-fitted metrics, there is generally agreement across large-scale model loss function and architecture trends across species and brain areas.
More specifically, studies commonly examine \emph{multiple} metrics rather than one, which can otherwise be susceptible to ``Goodharting''. 
These include both fitted metrics such as linear regression, along with non-fitted metrics (e.g. RSA, score function distributions, simpler-than-linear mappings, etc.), which found similar conclusions that matched linear regression results, across rodent medial entorhinal cortex~\citep[Fig. 3]{nayebi2021explaining}, macaque and human visual cortex (cf. \citet{khaligh2014deep} as the first of many such demonstrations), and human auditory cortex~\citep[Fig. S10]{tuckute2023many}.

However, one could make the argument that this general agreement among metrics is meaningless, since all the metrics we currently use could pick out ``functionally orthogonal'' features of neural data. 
One commonly referenced candidate for a functionally orthogonal feature is ``effective dimensionality''.
We note that it is debated whether this should even be a candidate, since it may be necessary even in the brain to maintain high effective dimensionality to support many different intelligent behaviors.
However, let us suppose for this discussion that it was undesirable if effective dimensionality unilaterally predicted brain-model alignment on a metric---is this true for common metrics?
This has largely been shown to \emph{not} be the case for linear regression or RSA, across brain areas and species, e.g. rodent medial entorhinal cortex where networks like the ``SimpleRNN''~\citep{Elman1990,nayebi2021explaining} have high neural predictivity but low participation ratio~\citep[Fig. 6]{schaeffer2022no}; human visual cortex and RSA~\citep[Fig. 5]{conwell2022can}; with current vision models and regression~\citep[Figs. SI5.10-SI5.12]{canatar2024spectral}; and human auditory cortex~\citep[Fig. S9]{tuckute2023many}.
These empirical findings are also supported by theoretical results in regression, such as in the recent spectral theory of linear regression of \citet{canatar2024spectral}, where there is an explicit model-brain alignment term ($Y^Tv_i$) of the model's principal components ($v_i$) being aligned with the neural data ($Y$).
Note the ``Universal Approximation Theorem''~\citep{cybenko1989approximation,hornik1989multilayer,hornik1991approximation} is inapplicable here for trivial reasons. 
Factors like out-of-distribution generalization -- since pre-training images often differ from images used for neural evaluation~\citep{yamins2014performance,schrimpf2020integrative}---along with finite layer widths and sample sizes, misalign the theorem with practice.

Instead, what we see emerging from these models often mirrors known features of neural processing in visual~\citep{yamins2014performance,khaligh2014deep,cadena2019deep,nayebi2018task,Zhuang2021,nayebi2021unsupervised,nayebi2022}, auditory~\citep{kell2018task,tuckute2023many}, motor~\citep{sussillo2015neural, michaels2020goal}, entorhinal~\citep{sorscher2019unified,nayebi2021explaining}, and other brain areas such as language~\citep{schrimpf2021neural}---such as Gabors, hierarchical structure, tuning properties, sparsity, and functional responses.
That is not to say the models are perfect (this is even evident from our metrics~\citep{schrimpf2020integrative}), but this is a reflection that the standard choices for the metric $\mathcal{M}$ are not pointing us \emph{away} from the brain. 
After all, the ``bread and butter'' of science is empirical, quantitative comparison, and this approach is standardizing this consideration to the brain sciences more formally.

Finally, from an oracle point of view, independent of \emph{any} model, we expect neural responses in the brain area of one animal to overlap in their encoding with the responses in the same brain area of another animal. 
This is a critical component of our NeuroAI Turing Test, where we strive for models that, at minimum, reach this criterion. 
In fact, models evaluated on many current metrics have not yet reached this consistency level, suggesting that we have not reached a point where models ``overfit'' to all available datasets (e.g. macaque dorsomedial frontal cortex and human physics prediction~\citep[Figs. 2B and 3B]{nayebi2023neural}, human auditory cortex~\citep[Fig. 2E]{tuckute2023many}, and human language areas~\citep[Fig. SI1]{schrimpf2021neural}).

\subsection{Finite Numbers of Models and Organisms}
\label{ss:against-finite}
Another criticism of the NeuroAI Turing Test is that the sets $\Delta_{\text{model}}$ and $\Delta_{\text{organism}}$ are inherently finite.
In other words, we rely on a finite sample of models and organisms (animals or human subjects), rather than the ``true'' continuous distribution.
Therefore, this leaves room for models which are not considered in any given test.

But this issue is part of \emph{every} empirical science.
Science has always been about comparing a finite number of models against data.
As a result, in science we can never claim a model (``theory'') \emph{is} the natural phenomenon, we can only rule models out and identify the model most consistent with measured phenomena~\citep{popper1934logik,lakatos1978methodology,doerig2023neuroconnectionist}.
 We can then test the remaining models of the brain with new experimental data and diverse similarity measures.
This process is very much in line with traditional scientific progress, in that we have hypotheses that we continually evaluate against data to see what is most consistent, up to the data’s ability to differentiate said hypotheses. 
If the model matches all data it has been tested against, then it cannot be rejected as a perfect model of the system. 
This is why it is important to test as stringently as possible---the point is to have more metrics, not less (or even just one)! 

Furthermore, the finite model hypotheses we pick are highly non-random, as they are the ones that exhibit intelligent behavior, rather than strawmen models or trivial direct fits to the data, making them non-trivially normative.
The more of these non-trivial hypotheses we have, the better positioned we are to identify theoretical invariances (if there are any).
In fact, we are even seeing now that functional models converge on similar representations across modalities~\citep{conwell2022can,huhposition,hosseini2024universality}, and that models that share invariances with other models are also more likely to share invariances with humans ~\citep[Fig. 8E,F]{feather2023model}.
Thus, identifying trends for invariances to later theoretically verify is strongly consistent with our measures of distributional equivalence for the NeuroAI Turing Test, since having more samples will naturally boost its accuracy.

It is also important to note that predictive measures go well beyond classic interventional controls in traditional psychological studies that only test a small number of situation-specific conditions. 
This shift to prediction as a higher-order readout of model-brain alignment is necessary because the hypothesis space for cognitive function in real-world environments is vast, with an enormous number of independent variables to consider that are not always guaranteed to be human-understandable.

Therefore, if such a deeper theory exists that governs the ``true'' distribution of invariant model representations that can perform intelligent tasks, then we can view what we are doing here as the necessary ``empirics gathering'' stage to strongly \emph{separate} hypotheses about what such a theory could even look like, through the study of successfully brain-predictive ANNs.
Almost every science begins this way, by collecting many empirical examples of a given phenomenon before identifying unifying principles (if any) responsible for the common trends that have been observed (e.g. the steam engine preceded thermodynamics).
This is especially the case since it is currently faster to build an ANN that is task-performant and predicts a brain area, than to theorize \emph{ab initio} about it and hope it explains the brain, as prior views of brain-likeness did (cf. \S\ref{ss:against-likeness}).

\subsection{Achievability of the NeuroAI Turing Test}
 \label{ss:against-achievable}
\textbf{One concern for model-brain comparisons is that our proposed NeuroAI Turing Test may be too lenient.}
We emphasize that for a finite sample of data, the NeuroAI Turing Test is a \emph{lower bound} on the ``true'' inter-organism consistency.
For instance, the inter-organism distance may be underestimated if available data are insufficient to capture the aspects of true distributional equivalence across organisms. In this case, many models may reach or surpass the inter-organism distance. Limited samples, topographical discrepancies between individuals, and noisy measurements can all degrade the dynamic range needed to discriminate genuinely brain-like models~\citep{dapello2022aligning}.

However, these are limitations of the datasets, not the NeuroAI Turing Test. By quantifying how inter-animal consistency changes as a function of the number of neurons recorded, the number of stimuli presented, or the number of repetitions for each stimulus, researchers can derive \emph{actionable} prescriptions for future data collection. For instance, ~\citet[Fig. S6B]{nayebi2021unsupervised} reported that expanding the stimulus set can increase inter-animal consistency and lead to a stronger dataset for brain-model comparisons.
Extrapolation analyses with existing data can further guide data collection efforts. For example, in mouse visual cortex, \citet[Fig. S5]{nayebi2021unsupervised} showed that under several choices of metric $\mathcal{M}$, a log-linear extrapolation analysis revealed that as the number of units increases, the inter-animal consistency approaches 1.0 more rapidly for Neuropixels data~\citep{de2020large}, than for calcium imaging data~\citep{Siegle2021}, indicating that Neuropixels as a data collection modality results in higher inter-animal consistencies. 
If the inter-animal alignment continues to improve with more extensive sampling of brain areas or expanded recording durations, it underscores the need to gather larger datasets. Conversely, if the inter-animal consistency begins to saturate, then the observed variability between organisms reflects individual differences in the population.

Importantly, if a model passes the NeuroAI Turing Test for a brain area under idealized data conditions, this may suggest that current neural datasets in these domains are not sufficiently complex to falsify the existing best models.
Just as the standard Turing Test involves multiple rounds of questioning to probe the depth of a machine's intelligence, the NeuroAI Turing Test should involve iterative refinement of the model and its evaluation on multiple new datasets with a diverse set of stimulus conditions and metrics. 
The ``Contravariance Principle'' of neural modeling~\citep{cao2024explanatory2} is helpful in this scenario: working in more complex experimental environments may make it easier to identify correct models by reducing susceptibility to overly simplistic data.

In other cases, models may surpass the brain-brain similarity, not because the data is impoverished but because the model representations reflect an implicit ``average'' brain rather than any single observed neural pattern. This phenomenon, akin to the ``wisdom of the crowd'' effect~\citep{stroop1932judgment}, has been observed in large language models (LLMs), where model-generated annotations often align more closely with humans than other individual human judgments~\citep{dillion2023can, trott2024large}. Similarly, in NeuroAI, models trained on diverse datasets may develop representations that reflect population-level patterns rather than idiosyncratic inter-organism variability. 
This could be particularly relevant for higher-level cognitive tasks, where individual brains may exhibit substantial variability (see Appendix \S\ref{ss:notions-variability} for discussion on an alternative test of models that includes such variability), but a model trained across many examples may develop representations that align with the central tendencies of human cognition and surpass the brain-brain similarity value.  

\textbf{Another concern for the NeuroAI Turing test is that it is impossibly stringent.} Under-achievability arises when all models fall short of inter-animal consistency---a situation that can occur in domains where the underlying neural or behavioral data are inherently complex. 
However, just as AI is making progress on the classic behavioral Turing Test, we believe our representational benchmark for model-brain similarity is also achievable. Further, representational convergence between computational models and brains has been achieved to a large extent in datasets and domains like the first 200 ms of macaque visual cortex~\citep{Majaj2015}, mouse visual cortex~\citep{nayebi2021unsupervised}, and rodent medial entorhinal cortex~\citep{nayebi2021explaining}. While it remains unattained in domains like audition~\citep{tuckute2023many}, language~\citep[Fig. SI12]{kauf2024lexical}, and physical understanding~\citep{nayebi2023neural,bear2021physion}, the rapid progress of modeling efforts suggests that models may reach representational convergence with the brain for these domains and datasets in the near future. 
Additionally, identifying challenging domains is critical.
Studying higher-level cognition can be difficult due to varying subject strategies, but rather than abandoning inter-subject comparisons, we advocate for metrics that extract shared features across trajectories, as in \citet{nayebi2021explaining}, where spatial averaging formed generalizable ``rate maps''.
Systematically tracking the gap between top models and inter-animal consistency reveals ``frontier'' datasets where models lag behind biological benchmarks.
These datasets offer fertile ground for targeted modeling and experimental improvements (see Appendix \S\ref{ss:notions-gap}), yielding more generalizable brain-like models.

\section{Conclusion}
In summary, the NeuroAI Turing Test provides a much-needed standard to bridge AI and neuroscience. 
Without it, AI systems may achieve high performance yet remain detached from any principled scientific notion of intelligence. 
Rather than treating the brain as an optimal blueprint, this framework positions it as a natural reference point---our only universally agreed-upon example of intelligence---while allowing for artificial systems to surpass it.
We call on the scientific community to adopt a rigorous common standard for evaluating models of intelligence, moving beyond loosely defined notions of biological similarity.

\section*{Acknowledgments}
A.N. thanks Leila Wehbe for helpful discussions, and the Burroughs Wellcome Fund CASI award for funding. N.A.R.M. is supported by the NIH Pathway to Independence Award (R00EY032603) from the National Eye Institute. J.F. is supported by the Flatiron Institute, a division of the Simons Foundation.

\section*{Impact Statement}
This work seeks to define a rigorous ``Turing Test'' not just for NeuroAI, but for the broader science and engineering of intelligence---establishing a systematic framework for evaluating models of intelligence in artificial and biological systems alike.
Rather than solely focusing on whether artificial models capture biological brain function, this framework provides a principled benchmark for assessing representational and behavioral alignment, helping to ground intelligence research across neuroscience, cognitive science, and AI.

By formalizing a standard for evaluating models, this work has the potential to improve the scientific rigor of computational neuroscience, advance cognitive science by clarifying functional and behavioral constraints on intelligence, and provide AI with a clearer benchmark for what constitutes a generalizable model of intelligence.
More broadly, it contributes to the growing effort to unify these disciplines into a predictive and functional science of intelligence---one that seeks \emph{algorithmic} principles abstracted from biological implementations and runnable on machines.

The societal impact of this research includes potential applications in medicine, brain-computer interfaces (BCIs), and neurotechnology, which we briefly discuss in this article. 
A clearer framework for modeling intelligence may aid in understanding neural computations relevant to neurological disorders or cognitive interventions. 
However, as with all work at the intersection of neuroscience and AI, ethical considerations arise regarding data privacy, informed consent in neural recording, and the potential misuse of predictive models of brain activity.

Ultimately, this work is conceptual and methodological, with no immediate ethical risks beyond those already inherent to advancing AI and neuroscience. 
Nonetheless, we emphasize the importance of continued discussion on the responsible development and application of intelligence research, especially the responsible use of AI in understanding brain function.

\textbf{We encourage a dialogue: Please feel free to contact us if you have any questions or suggestions!}

\bibliography{references}
\bibliographystyle{icml2025}

\newpage
\appendix
\onecolumn
\section{Inter-Subject Noise Correction Derivation}
\label{sec:methods-interanimal}
Herein we describe how the metric $\mathcal{M}$ should correct for noise if there is trial-to-trial variability.
This is unified and adapted from~\citet{nayebi2021unsupervised,nayebi2021explaining,nayebi2023neural}.

If you prefer to skip the derivation, for correlation-based metrics this yields the following model-to-data mapping correction: Let $L$ be the set of model layers, let $r^{\ell}$ be the set of model responses at model layer $\ell \in L$, $M$ be the mapping, and let $\mathrm{s}$ be the trial-averaged pseudo-population response.

\begin{equation}\label{modelcon}
\begin{split}
\max_{\ell\in L}\median\bigoplus_{\animalB \in \mathcal{A}} \left\langle\dfrac{\corr\left(M\left(r^{\ell}_{\text{train}};\sfBtrain\right)_{\test}, \ssBtest\right)}{\sqrt{\widetilde{\corr}\left(M\left(r^{\ell}_{\text{train}};\sfBtrain\right)_{\test}, M\left(r^{\ell}_{\text{train}};\ssBtrain\right)_{\test}\right) \times \widetilde{\corr}\left(\sfBtest, \ssBtest\right)}}\right\rangle,
\end{split}
\end{equation}
where the average is taken over bootstrapped split-half trials, $\oplus$ denotes concatenation of units across animals or subjects $\animalB \in \mathcal{A}$ followed by the median value across units, and $\corr(\cdot,\cdot)$ denotes the Pearson correlation of the two quantities.
$\widetilde{\corr}(\cdot, \cdot)$ denotes the Spearman-Brown corrected value of the original quantity (cf. \S\ref{ss:methods-interanimal-spearman-brown}).
We derive the analogous correction for RSA in \S\ref{ss:methods-interanimal-rsa}.

The above correction in \eqref{modelcon} is fully implemented in the 
\texttt{brainmodel\_utils} package (\url{https://github.com/neuroagents-lab/brainmodel_utils}), 
specifically in the \texttt{get\_linregress\_consistency} function. 
This function can be imported as follows:

\begin{lstlisting}
from brainmodel_utils.metrics.consistency import get_linregress_consistency
\end{lstlisting}

The \texttt{r\_xy\_n\_sb} value returned by this function corresponds to the ratio in \eqref{modelcon}.
Refer to the
\href{https://github.com/neuroagents-lab/brainmodel_utils#readme}{README} and the function docstring for usage details across a range of linearly regressed and non-regressed (e.g. RSA) metrics.

\subsection{Single Subject Pair}
\label{ss:methods-interanimal-pair}
Suppose we have neural responses from two animals (or subjects) $\animalA$ and $\animalB$.
Let $\mathrm{t}_i^p$ be the vector of true responses (either at a given time bin or averaged across a set of time bins) of animal $p \in \mathcal{A} = \{\animalA,\animalB,\dots\}$ on stimulus set $i \in \{\train, \test\}$.
Of course, we only receive noisy observations of $\mathrm{t}_i^p$, so let $\mathrm{s}_{j,i}^p$ be the $j$th set of $n$ trials of $\mathrm{t}_i^p$.
Finally, let $M(x;y)_i$ be the predictions of a mapping $M$ (e.g., PLS) when trained on input $x$ to match output $y$ and tested on stimulus set $i$.
For example, $M\left(\trueA;\trueB\right)_{\test}$ is the prediction of mapping $M$ on the test set stimuli trained to match the true neural responses of animal $\animalB$ given, as input, the true neural responses of animal $\animalA$ on the train set stimuli.
Similarly, $M\left(\sfAtrain;\sfBtrain\right)_{\test}$ is the prediction of mapping $M$ on the test set stimuli trained to match the trial-average of noisy sample 1 on the train set stimuli of animal $\animalB$ given, as input, the trial-average of noisy sample 1 on the train set stimuli of animal $\animalA$.

With these definitions in hand, the inter-animal mapping consistency from animal $\animalA$ to animal $\animalB$ corresponds to the following ``true'' quantity to be estimated by $\Mest$ in the limit of infinite trials:
\begin{equation}\label{interancontrue}
\mathcal{M}_{\text{true}} := \corr\left(M\left(\trueA;\trueB\right)_{\test}, \trueBtest\right),
\end{equation}
where $\corr(\cdot, \cdot)$ is the Pearson correlation across a stimulus set.
In what follows, we will argue that Eq~\eqref{interancontrue} can be approximated with the following ratio of measurable quantities, where we split in half and average the noisy trial observations, indexed by 1 and by 2:
\begin{equation}\label{interancon}
\begin{split}
& \Mtrue := \corr\left(M\left(\trueA;\trueB\right)_{\test}, \trueBtest\right) \\
& \sim \Mest := \dfrac{\corr\left(M\left(\sfAtrain;\sfBtrain\right)_{\test}, \ssBtest\right)}{\sqrt{\corr\left(M\left(\sfAtrain;\sfBtrain\right)_{\test}, M\left(\ssAtrain;\ssBtrain\right)_{\test}\right) \times \corr\left(\sfBtest, \ssBtest\right)}}.
\end{split}
\end{equation}
In words, the inter-animal consistency (i.e., the quantity on the left side of Eq~\eqref{interancon}) corresponds to the predictivity of the mapping on the test set stimuli from animal $\animalA$ to animal $\animalB$ on two different (averaged) halves of noisy trials (i.e., the numerator on the right side of Eq~\eqref{interancon}), corrected by the square root of the mapping reliability on animal $\animalA$'s responses to the test set stimuli on two different halves of noisy trials multiplied by the internal consistency of animal $\animalB$.

We justify the approximation in Eq~\eqref{interancon} by gradually replacing the true quantities ($\mathrm{t}$) by their measurable estimates ($\mathrm{s}$), starting from the original quantity in Eq~\eqref{interancontrue}.
First, we make the approximation that:
\begin{equation}\label{step1}
\corr\left(M\left(\trueA;\trueB\right)_{\test}, \ssBtest\right) \sim \corr\left(M\left(\trueA;\trueB\right)_{\test}, \trueBtest\right) \times \corr\left(\trueBtest, \ssBtest\right),
\end{equation}
by the transitivity of very positive correlations. 
Namely, in scenarios where correlations are very close to 1, a form of transitivity holds, meaning if variable $A$ is highly correlated with variable $B$, and variable $B$ with variable $C$, then variable $A$ is also highly correlated with variable $C$. 
This is the desired situation, as low or negative correlations indicate neurons that are not self-consistent.
Moreover, calculating certain metrics in these cases can result in undefined values due to operations like taking the square root of a negative number. 
Assuming high correlations is reasonable, especially when the number of stimuli is large.
Next, by transitivity and normality assumptions in the structure of the noisy estimates and since the number of trials ($n$) between the two sets is the same, we have that:
\begin{align}\label{step2}
\corr\left(\sfBtest, \ssBtest\right) &\sim \corr\left(\sfBtest, \trueBtest\right) \times \corr\left(\trueBtest, \ssBtest\right) \nonumber \\
&\sim \corr\left(\trueBtest, \ssBtest\right)^2.
\end{align}
In words, Eq~\eqref{step2} states that the correlation between the average of two sets of noisy observations of $n$ trials each is approximately the square of the correlation between the true value and average of one set of $n$ noisy trials.
Therefore, combining Eq~\eqref{step1} and Eq~\eqref{step2}, it follows that:
\begin{equation}\label{lemma1}
\corr\left(M\left(\trueA;\trueB\right)_{\test}, \trueBtest\right) \sim \dfrac{\corr\left(M\left(\trueA;\trueB\right)_{\test}, \ssBtest\right)}{\sqrt{\corr\left(\sfBtest, \ssBtest\right)}}.
\end{equation}

From the right side of Eq~\eqref{lemma1}, we can see that we have removed $\trueBtest$, but we still need to remove the $M\left(\trueA;\trueB\right)_{\test}$ term, as this term still contains unmeasurable (i.e., true) quantities.
We apply the same two steps, described above, by analogy, though these approximations may not always be true (they are, however, true for Gaussian noise):
\begin{equation*}
\begin{split}
\corr\left(M\left(\sfAtrain;\sfBtrain\right)_{\test}, \ssBtest\right) & \sim \corr\left(\ssBtest, M\left(\trueA;\trueB\right)_{\test}\right) \\
& \times \corr\left(M\left(\trueA;\trueB\right)_{\test}, M\left(\sfAtrain;\sfBtrain\right)_{\test}\right)
\end{split}
\end{equation*}
\begin{equation*}
\begin{split}
& \corr\left(M\left(\sfAtrain;\sfBtrain\right)_{\test}, M\left(\ssAtrain;\ssBtrain\right)_{\test}\right) \\
& \sim \corr\left(M\left(\sfAtrain;\sfBtrain\right)_{\test}, M\left(\trueA;\trueB\right)_{\test}\right)^2,
\end{split}
\end{equation*}
which taken together implies the following:
\begin{equation}\label{lemma2}
\corr\left(M\left(\trueA;\trueB\right)_{\test}, \ssBtest\right) \sim \dfrac{\corr\left(M\left(\sfAtrain;\sfBtrain\right)_{\test}, \ssBtest\right)}{\sqrt{\corr\left(M\left(\sfAtrain;\sfBtrain\right)_{\test}, M\left(\ssAtrain;\ssBtrain\right)_{\test}\right)}}.
\end{equation}
Eq~\eqref{lemma1} and Eq~\eqref{lemma2} together imply the final estimated quantity given in Eq~\eqref{interancon}.

\subsection{Multiple Subject Pairs}
\label{ss:methods-interanimal-multiple}
For multiple animals, we consider the average of the true quantity for each target in $\animalB$ in Eq~\eqref{interancontrue} across source animals $\animalA$ in the ordered pair $(\animalA,\animalB)$ of animals $\animalA$ and $\animalB$:
\begin{equation*}\label{multipleinterancon}
\begin{split}
&\Mtrue := \left\langle \corr\left(M\left(\trueA;\trueB\right)_{\test}, \trueBtest\right)\right\rangle_{\animalA \in \mathcal{A}: (\animalA,\animalB)\in \mathcal{A}\times\mathcal{A}} \\
& \sim \Mest := \left\langle\dfrac{\corr\left(M\left(\sfAtrain;\sfBtrain\right)_{\test}, \ssBtest\right)}{\sqrt{\widetilde{\corr}\left(M\left(\sfAtrain;\sfBtrain\right)_{\test}, M\left(\ssAtrain;\ssBtrain\right)_{\test}\right) \times \widetilde{\corr}\left(\sfBtest, \ssBtest\right)}}\right\rangle_{\animalA \in \mathcal{A}: (\animalA,\animalB)\in \mathcal{A}\times\mathcal{A}}.
\end{split}
\end{equation*}
We also bootstrap across trials, and have multiple train/test splits, in which case the average on the right hand side of the equation includes averages across these as well.

Note that each neuron in our analysis will have this single average value associated with it when \emph{it} was a target animal ($\animalB$), averaged over source animals/subsampled source neurons, bootstrapped trials, and train/test splits.
This yields a vector of these average values, which we can take median and standard error of the mean (s.e.m.) over, as we do with standard explained variance metrics.

\subsection{RSA}
\label{ss:methods-interanimal-rsa}
We can extend the above derivations to other commonly used metrics for comparing representations that involve correlation.
Since $\rsa(x,y) \coloneqq \corr(\rdm(x), \rdm(y))$, then the corresponding quantity in Eq~\eqref{interancon} analogously (by transitivity of maximally positive correlations) becomes:
\begin{equation}\label{rsainterancon}
\begin{split}
&\Mtrue := \left\langle\rsa\left(M\left(\trueA;\trueB\right)_{\test}, \trueBtest\right)\right\rangle_{\animalA \in \mathcal{A}: (\animalA,\animalB)\in \mathcal{A}\times\mathcal{A}} \\
& \sim \Mest := \left\langle\dfrac{\rsa\left(M\left(\sfAtrain;\sfBtrain\right)_{\test}, \ssBtest\right)}{\sqrt{\widetilde{\rsa}\left(M\left(\sfAtrain;\sfBtrain\right)_{\test}, M\left(\ssAtrain;\ssBtrain\right)_{\test}\right) \times \widetilde{\rsa}\left(\sfBtest, \ssBtest\right)}}\right\rangle_{\animalA \in \mathcal{A}: (\animalA,\animalB)\in \mathcal{A}\times\mathcal{A}}.
\end{split}
\end{equation}

Note that in this case, each \emph{animal} (rather than neuron) in our analysis will have this single average value associated with it when \emph{it} was a target animal ($\animalB$) (since RSA is computed over images and neurons), where the average is over source animals/subsampled source neurons, bootstrapped trials, and train/test splits.
This yields a vector of these average values, which we can take median and s.e.m. over, across animals $\animalB \in \mathcal{A}$.

For RSA, we can use the identity mapping (since RSA is computed over neurons as well, the number of neurons between source and target animal can be different to compare them with the identity mapping). 
As parameters are not fit, we can choose $\train = \test$, so that Eq~\eqref{rsainterancon} becomes:
\begin{equation}\label{rsainteranconid}
\Mtrue := \left\langle\rsa\left(\trueAid,\trueBid\right)\right\rangle_{\animalA \in \mathcal{A}: (\animalA,\animalB)\in \mathcal{A}\times\mathcal{A}} \sim \Mest := \left\langle\dfrac{\rsa\left(\sfAid, \ssBid\right)}{\sqrt{\widetilde{\rsa}\left(\sfAid, \ssAid\right) \times \widetilde{\rsa}\left(\sfBid, \ssBid\right)}}\right\rangle_{\animalA \in \mathcal{A}: (\animalA,\animalB)\in \mathcal{A}\times\mathcal{A}}.
\end{equation}

\subsection{Pooled Source Animal}
\label{ss:methods-interanimal-holdouts}
Often times, we may not have enough neurons per animal to ensure that the estimated inter-animal consistency in our data closely matches the ``true'' inter-animal consistency.
In order to address this issue, we holdout one animal at a time and compare it to the pseudo-population aggregated across units from the remaining animals, as opposed to computing the consistencies in a pairwise fashion.
Thus, $\animalB$ is still the target heldout animal as in the pairwise case, but now the average over $\animalA$ is over a sole ``pooled'' source animal constructed from the pseudo-population of the remaining animals.

Pooling data across subjects to create larger pseudopopulations is a common practice~\citep{rust2020understanding}, and helps researchers better isolate core representational principles that are conserved across individuals when data collection modalities limit the number of collected neurons per session.

\subsection{Spearman-Brown Correction}
\label{ss:methods-interanimal-spearman-brown}
The Spearman-Brown correction can be applied to each of the terms in the denominator individually, as they are each correlations of observations from half the trials of the \emph{same} underlying process to itself (unlike the numerator). Namely,
\begin{equation*}
\widetilde{\corr}\left(X,Y\right) \coloneqq \frac{2\corr\left(X,Y\right)}{1 + \corr\left(X,Y\right)}.
\end{equation*}
Analogously, since $\rsa(X,Y) \coloneqq \corr(\rdm(x), \rdm(y))$, then we define
\begin{align*}
\widetilde{\rsa}\left(X,Y\right) &\coloneqq \widetilde{\corr}(\rdm(x), \rdm(y)) \\
    &= \frac{2\rsa\left(X,Y\right)}{1 + \rsa\left(X,Y\right)}.
\end{align*}

\section{Notions of Brain-Likeness}
\label{sec:notions}
In developing artificial intelligence that mirrors the biological brain, the concept of brain-likeness has been interpreted in various ways. Here, we explore the most common notions of brain-likeness (cf. Table~\ref{tab:brain_likeness} for a summary):

\paragraph{Brain-Likeness in structural components} refers to how closely a model’s internal architecture---such as layer organization, connectivity patterns, and computational units---resembles that of the brain. Modern convolutional neural networks (CNNs) exemplify this by integrating key neural computation principles like nonlinear transduction, divisive normalization, and max-based pooling~\citep{yamins2016using}. These design choices are directly inspired by electrophysiological studies of the mammalian visual cortex, where simple and complex cells in area V1 filter and pool visual inputs to extract features at different scales~\citep{Hubel1962}. Similarly, recurrent neural networks (RNNs) capture temporal dependencies through recurrent connectivity, paralleling biological circuits where feedback connections are essential for maintaining and integrating information over time, as seen in the prefrontal cortex supporting working memory and sequential processing~\citep{goldman1995cellular}. Spiking neural networks (SNNs) strive for even greater structural similarity by incorporating neuron-like spiking activity and temporal dynamics, reflecting biological processes such as synaptic integration and spike-timing-dependent plasticity~\citep{gerstner2002spiking}. Additionally, modular architectures in AI, where different modules specialize in tasks like memory, control, or perception, are inspired by the brain’s division of labor across distinct but interacting subsystems, such as the hippocampus for memory and the prefrontal cortex for decision-making~\citep{anderson2004integrated}. This modularity facilitates domain-specific processing in a manner that mirrors biological brain organization.

\paragraph{Brain-Likeness in inductive biases} involves the inherent assumptions or constraints within a model’s design that reflect principles believed to govern neural computation. For example, sparse coding in models like Sparse Autoencoders~\citep{ng2011sparse} utilizes sparse activation patterns to prioritize energy-efficient information encoding, mirroring the brain’s efficient neural representation~\citep{olshausen2004sparse}. Equivariance in CNNs allows these networks to recognize objects regardless of their orientation or position, reflecting the brain’s transformational invariance in object recognition~\citep{yamins2016using}. These inductive biases help create internal representations that are more akin to those found in the brain, although models with such biases may still process information differently from biological systems.

\textbf{Brain-likeness in training} examines whether the learning processes of models resemble the experience-driven development that shapes biological brains. Models may employ learning paradigms that mirror the brain’s ability to develop through experience and interaction with the environment. For example, self-supervised learning, as in contrastive learning models like SimCLR~\citep{chen2020simple}, parallels the brain’s ability to learn from unlabelled sensory inputs. Curriculum learning~\citep{bengio2009curriculum} involves progressively tackling more complex tasks, mirroring the staged learning observed in human development, where simpler skills are acquired before more complex ones. Furthermore, embodied agents that interact with their environment to acquire knowledge reflect the brain's dynamic learning through active engagement and adaptation to changing conditions~\citep{pfeifer2006body}. Similarly, curiosity-driven learning~\citep{pathak2017curiosity} encourages models to explore by generating intrinsic rewards based on prediction errors in learned feature spaces, mirroring how biological agents seek out novel, informative experiences to guide learning in the absence of explicit external rewards. Regularization techniques in deep learning can also be drawn from biological principles. Dropout, which randomly deactivates a subset of units during training, was motivated by the stochastic nature of biological neurons, which exhibit Poisson-like firing variability~\citep{hinton2012improving}.

\paragraph{Brain-likeness in computational principles} 
A model can be considered brain-like if its computational or theoretical foundations align with well-established principles of neural computation. 
For example, the most notable of these include classic ideas of predictive coding~\citep{rao1999predictive}, sparse coding~\citep{olshausen1996emergence}, energy efficiency~\citep{laughlin2001energy}, and redundancy reduction~\citep{barlow1961possible}.

\begin{table*}[t]
    \centering
    \caption{Comparative overview of alternate notions of ``brain-likeness''}
    \label{tab:brain_likeness}
    \begin{tabular}{>{\bfseries\raggedright\arraybackslash}m{2.5cm} >{\raggedright\arraybackslash}m{4.5cm} >{\raggedright\arraybackslash}m{5cm} >{\raggedright\arraybackslash}m{4cm}}
        \toprule
        \textbf{Notion of Brain-Likeness} & \textbf{Description} & \textbf{Key Examples} & \textbf{References} \\
        \midrule
        \textbf{Structural Components} & Resemblance in architecture and connectivity patterns & Convolutional Neural Networks (CNNs), Recurrent Neural Networks (RNNs), Spiking Neural Networks (SNNs), Modular Architectures & Yamins \& DiCarlo (2016); Hubel \& Wiesel (1962) \\
        
        \textbf{Inductive Biases} & Inherent assumptions or constraints in model design & Sparse Autoencoders, Equivariance in CNNs & Olshausen \& Field (1996); Hinton et al. (2012) \\
        
        \textbf{Training Paradigms} & Learning processes that mirror biological development & Self-Supervised Learning (SimCLR), Curriculum Learning, Curiosity-Driven Learning & Chen et al. (2020); Pathak et al. (2017) \\
        
        \textbf{Computational Principles} & Theoretical frameworks guiding neural computation & Predictive Coding, Energy Efficiency, Redundancy Reduction & Rao \& Ballard (1999); Laughlin (2001); Barlow (1961) \\
        \bottomrule
    \end{tabular}
\end{table*}

\subsection{Matching model variability to animal variability}
\label{ss:notions-variability}
A more stringent future direction for model development involves ensuring that the variability observed in models aligns with the variability observed in biological systems. If a model posits a mechanism of variability---such as differences arising from random initial states (e.g., random seeds), stochastic training processes, or architectural variations---it should exhibit a distribution of representational alignments that matches the inter-animal variability seen in neural data. For example, if the alignment scores between a model and neural data vary significantly across different random initializations (or any assumed mechanism of variability), this variability should resemble the natural variability observed across individuals within a species. This requirement ensures that models not only capture the shared computational principles of the brain but also reflect the inherent diversity and noise present in biological systems. By incorporating this constraint, we can develop models that are not only brain-like in their representations but also in their variability, further bridging the gap between artificial and biological intelligence.

\section{Modeling efforts for closing the gap in the NeuroAI Turing Test}
\label{ss:notions-gap}
Despite significant progress in NeuroAI, across various domains, current models remain far from achieving representations that are indistinguishable from those found in the brain. These discrepancies can be broadly categorized into architectural, training, and data-related gaps, each presenting unique challenges that require novel solutions. Additionally, enhancing models to exhibit more brain-like behaviors---such as increased robustness, adaptability, and generalization---can further contribute to bridging the gap in achieving brain-like representations.
\paragraph{Architectural gaps}
The architecture of most artificial neural networks diverges significantly from the brain’s structure, but selectively incorporating certain biological features could help achieve more brain-like representations. Recurrent dynamics, for instance, enable feedback loops and temporal processing~\citep{goldman1995cellular}, allowing models to maintain context, integrate information over time, and develop richer internal states---capabilities that are often limited in feedforward architectures. Modularity, another key feature, mirrors the brain’s organization into specialized, functionally distinct regions that interact hierarchically and in parallel. This could enhance models’ ability to decompose tasks, generalize across domains, and scale efficiently. Additionally, the heterogeneity of neuron types, such as excitatory and inhibitory neurons, plays a critical role in balancing and regulating network activity, which could improve the stability and robustness of artificial systems~\citep{rubin2017balanced}. Another promising feature is sparsity~\citep{olshausen2004sparse}, which reflects the brain’s efficient use of sparse connectivity and activation patterns, reducing computational costs while enabling more flexible and interpretable representations. By thoughtfully integrating these biologically inspired features---recurrent dynamics, modularity, heterogeneity of neuron types, and sparsity---we can develop models that not only perform tasks more effectively but also exhibit internal representations that more closely resemble those of the brain.

\paragraph{Training gaps}
The training paradigms for artificial neural networks often lack key elements that are central to how the brain learns. Unlike the brain, which learns through multimodal training---integrating information from vision, sound, touch, and other senses---many AI models are trained on unimodal datasets, limiting their ability to develop rich representations that are essential for building accurate world models. Additionally, the brain’s learning is deeply rooted in embodiment~\citep{pfeifer2006body}, where sensory and motor experiences are tightly coupled, enabling agents to interact with and learn from their environment in a grounded, context-dependent manner. This is closely tied to agency~\citep{baldassarre2013intrinsically}, the capacity to take actions and observe their consequences, which drives curiosity, exploration, and goal-directed behavior. Incorporating these principles---multimodal training, embodiment, and agency---into AI systems could foster more robust, adaptive, and generalizable learning, possibly also bringing models internal representations closer to the brain.

\paragraph{Data gaps}
A significant challenge in developing brain-like representations lies in the limitations of the data used to train artificial neural networks. Current datasets, such as ImageNet~\citep{Deng2009} for vision and AudioSet~\citep{gemmeke2017audio} for audio, often fail to match the vast and continuous stream of sensory input the brain processes over a lifetime. ImageNet, for instance, contains static, curated images that lack the temporal dynamics and contextual richness of real-world visual experiences, limiting the ability of models to learn spatiotemporal relationships and contextual dependencies. Similarly, AudioSet, though large, is limited in its coverage of natural auditory scenes and their interactions with other modalities, restricting the development of multimodal representations that are critical for understanding complex environments. These datasets also lack diversity and naturalism, as they are often domain-specific and collected in controlled settings, failing to capture the noisy, unstructured, and interactive qualities of real-world experiences. Newer datasets like Ego4D~\citep{grauman2022ego4d}, which focuses on egocentric, multimodal video data captured from wearable cameras, aim to address these gaps by providing large-scale, diverse, and naturalistic datasets that reflect the dynamic, first-person perspective of human interaction with the world. By training on such data, models can develop more brain-like internal representations---rich, context-aware, and multimodal---that mirror the brain’s ability to process and integrate complex, real-world information

The gap in achieving brain-like representations in NeuroAI reflects limitations across architecture, training paradigms, and data. While addressing these gaps individually is crucial, they are deeply interconnected: improvements in architecture and training can benefit from better data, and behavioral alignment can provide constraints that shape internal representations to be more brain-like. Closing the gap for brain-like behaviors, including robustness, adaptability, and generalization, is also likely to drive corresponding improvements in representational alignment.

\section{Metrics for representational comparisons} 
\label{ss:notions-metrics}
No single metric can fully capture the complexity of brain-like representations. Therefore, a combination of complementary metrics that probe different dimensions of neural alignment should be used to provide a comprehensive evaluation. Metrics for evaluating model-brain alignment can be broadly divided into two main categories: alignment-based measures and representational similarity matrix (RSM)-based measures. Alignment-based measures quantify the distance or similarity between model and neural representations after aligning their dimensions, while RSM-based measures compare stimulus-by-stimulus similarity matrices derived from model and neural data. These categories can be further subdivided based on symmetry, scope of similarity (global vs. local), and level of alignment (population vs. unit-level).
Alignment-based measures include methods like linear predictivity~\citep{yamins2014performance}, canonical correlation analysis (CCA)~\citep{hotelling1992relations}, Procrustes distance~\citep{williams2021generalized}, soft matching~\citep{khosla2024soft}, and pairwise matching~\citep{khosla2024privileged, li2015convergent}. These methods differ in whether they quantify similarity at the population level (e.g., overall information content, as in CCA or linear predictivity) or the unit level (e.g., correspondence between individual model and neural units, as in soft or pairwise matching). They also differ in symmetry: some are symmetric (e.g., CCA, Procrustes, soft matching), treating both representations equally, while others are asymmetric (e.g., linear predictivity, pairwise matching), reflecting directional relationships.
RSM-based measures, such as representational similarity analysis (RSA)~\citep{kriegeskorte2008representational}, centered kernel alignment (CKA)~\citep{kornblith2019similarity}, and mutual $k$-nearest neighbors (mutual $k$-NN)~\citep{huhposition}, focus on comparing the similarity structure of model and neural representations. These methods can be distinguished by their scope of similarity: global metrics like RSA and CKA use all samples to compute distances, capturing the overall geometry of the representational space, while local metrics like mutual $k$-NN focus on neighborhood consistency, evaluating whether individual stimuli have similar neighboring points across the two representational spaces.

Each metric captures a distinct dimension of similarity and together, they provide complementary insights into the nature of alignment between models and brains (cf. Table~\ref{tab:metrics} for a summary). For example, global metrics like RSA and CKA reveal whether the model captures the overall geometry of neural representations, while local metrics like mutual $k$-NN assess whether local neighborhoods of a stimulus are similar across the model and brain representations. Alignment-based measures like linear predictivity and CCA evaluate how well the model’s information content aligns with neural data, while unit-level measures like soft matching test whether individual model units correspond to specific neural units. A model might close the gap in terms of overall information content (e.g., as measured by linear predictivity or CCA) but fail to replicate unit-level responses that are consistent across animals (e.g., as measured by soft matching or pairwise matching). This discrepancy would indicate that while the model captures the global structure of neural representations, important fine-grained details---such as the specific tuning curves or response properties of individual units---remain unexplained. Such findings can prompt novel model development efforts, particularly those focused on producing the correct axes of neural representations (e.g., tuning curves, receptive fields) that are conserved across individuals. By systematically evaluating models across multiple metrics, we can identify which dimensions of alignment are well-matched and which require further refinement, ultimately driving the development of models that more fully emulate the brain’s representational properties.

\begin{table}[t]
\centering
\caption{Categorization of common metrics for model-brain alignment. 
Symmetry and scope are also indicated for each metric.}
\label{tab:metrics}
\resizebox{0.85\columnwidth}{!}{
\begin{tabular}{ll lll}
\toprule
\textbf{Category}         & \textbf{Subcategory}         & \textbf{Metric}        & \textbf{Symmetry}  & \textbf{Scope}       \\
\midrule
\multirow{5}{*}{\textbf{Alignment-Based}}    
                          & \multirow{3}{*}{Population-Level Similarity} 
                          & Linear Predictivity         & Asymmetric   & Global  \\
                          &                              & CCA                     & Symmetric    & Global  \\
                          &                              & Procrustes              & Symmetric    & Global  \\
                          & \multirow{2}{*}{Unit-Level Similarity}   
                          & Soft Matching              & Symmetric    & Global  \\
                          &                              & Pairwise Matching       & Asymmetric   & Global  \\
\midrule
\multirow{3}{*}{\textbf{RSM-Based}}  
                          & \multirow{2}{*}{Global Similarity}  
                          & RSA                         & Symmetric    & Global  \\
                          &                              & CKA                     & Symmetric    & Global  \\
                          & Local Similarity            & Mutual $k$-NN           & Symmetric    & Local   \\
\bottomrule
\end{tabular}
}
\end{table}

\end{document}